\documentclass[%
 reprint,
superscriptaddress,
 amsmath,amssymb,
 aps,
]{revtex4-1}

\usepackage{graphicx}
\usepackage{dcolumn}
\usepackage{bm}

\makeatletter
\newcommand{\ccaption}[2][]{%
	\begingroup%
	\renewcommand{\@caption@fignum@sep}{ (color online). }%
	\caption[#1]{#2}%
	\endgroup%
}
\makeatother

\begin{document}

\preprint{APS/123-QED}

\title{Microscopic treatment of energy partition in fission}

\author{M. Albertsson}
\affiliation{Mathematical Physics, Lund University, S-221\,00 Lund, Sweden}
\author{B.G. Carlsson}
\affiliation{Mathematical Physics, Lund University, S-221\,00 Lund, Sweden}
\author{T. D{\o}ssing}
\affiliation{Niels Bohr Insitute, DK-2100 Copenhagen {\O}, Denmark}
\author{P. M{\"o}ller}
\affiliation{Mathematical Physics, Lund University, S-221\,00 Lund, Sweden}
\affiliation{Theoretical Division, Los Alamos National Laboratory, 
			 Los Alamos, New Mexico 87545, USA}
\affiliation{P.\ Moller Scientific Computing and Graphics, Inc.,
		 Los Alamos, New Mexico 87544, USA}
\author{J. Randrup}
\affiliation{Nuclear Science Division, Lawrence Berkeley National Laboratory, Berkeley, California 94720, USA}
\author{S. {\AA}berg}
\affiliation{Mathematical Physics, Lund University, S-221\,00 Lund, Sweden}

\date{\today}

\begin{abstract}
The transformation of an atomic nucleus into two excited fission fragments 
is modeled as a strongly damped evolution of the nuclear shape,
until scission occurs at a small critical neck radius, at which point 
the mass, charge, and shape of each fragment are extracted.
The available excitation energy then is divided statistically
on the basis of the microscopic level densities.
This approach takes account of the important (and energy-dependent) 
finite-size effects.
After the fragments have been fully accelerated 
and their shapes have relaxed to their equilibrium form, 
they undergo sequential neutron evaporation.
The dependence of the resulting mean neutron multiplicity 
on the fragment mass, $\bar{\nu}(A)$, 
including the dependence on the initial excitation energy
of the fissioning compound nucleus,
is in good agreement with the observed behavior,
as demonstrated here for $^{235}$U(n,f).
\end{abstract}

\maketitle
\raggedbottom

Even 80 years after its discovery \cite{Hahn1939,Meitner1939}, 
nuclear fission remains a fertile topic 
for experimental and theoretical research
\cite{AndreyevRPP81,SchmidtRPP81,SchunckRPP79,TalouEPJA54}
and improvements in instrumentation,
modeling, and computation have enabled a renaissance in the field.

In their seminal paper \cite{BohrPR56},
Bohr and Wheeler described fission as an evolution of the nuclear shape 
subject to both conservative forces from the potential energy of deformation
and dissipative forces resulting from the coupling to the residual system.
This conceptually simple picture suggests that the shape dynamics
can be regarded as a Brownian process, 
as pioneered by Kramers \cite{KramersPhysica7}.
In the idealized limit of strong dissipation,
the shape evolution can then be simulated by a random walk 
 \cite{RandrupPRL106,RandrupPRC84}
on the multi-dimensional potential-energy surface,
from near the ground state shape, across the barrier region,
until the system divides into two fragments at scission.

In a recent study \cite{WardPRC95}, 
it was shown that the use of shape-dependent microscopic level densities 
for guiding the Brownian shape evolution
provides a consistent (and parameter-free) framework 
for calculating the energy-dependent fission-fragment mass distribution.

We develop that approach further by partitioning 
the available excitation energy at scission
between the two nascent fragments
based on their microscopic level densities.
The division of the available energy between the two fragments 
has long been puzzling because it appears to differ from 
expectations based on their masses.
However, it was recently pointed out \cite{SchmidtPRL104}
that previous treatments \cite{MadlandNSE81,LemairePRC72,KornilovNPA789}
employed the simplified Fermi-gas (Bethe) formula \cite{BethePR50}
which may not be accurate at low energies
where structure effects tend to be significant.

We demonstrate that a consistent use of 
the appropriate microscopic level densities
in the distorted pre-fragments at scission 
leads to a remarkably good reproduction of the experimental data.

In our study, we generate and analyze a large number 
of scission configurations (typically $10^6$)
for the compound system $^{236}{\rm U}^*$ 
having a specified initial excitation energy $E_0^*$.
For this task, 
we employ the Brownian shape evolution method \cite{RandrupPRL106},
performing Metropolis walks on the potential-energy surface tabulated 
for the three-quadratic-surfaces shape family \cite{MollerPRC79}.
These shapes \cite{NixNPA130} are characterized by five parameters: 
the overall elongation given by the quadrupole moment $Q$,
the radius $c$ of the hyperbolic neck between the two spheroidal 
end sections which have deformations $\varepsilon_1$ and $\varepsilon_2$,
and the mass asymmetry $\alpha$.
Each Metropolis walk is started near the second minimum and continued 
across and beyond the outer barrier until the neck radius $c$ 
has become smaller than the specified critical value $c_0$=1.5 fm.

At scission, the value of the asymmetry parameter $\alpha$ determines 
the mass numbers of the nascent heavy and light fragments, $A_H$ and $A_L$.
The associated charge numbers, $Z_H$ and $Z_L$, are selected as 
those values that best preserve the $N$:$Z$ ratio.
(For simplicity, we consider only divisions into even-even fragments
in this first exploratory study.)
Furthermore,
the $\varepsilon$ parameters give the spheroidal deformations
of the nascent fragments, $\varepsilon_i^{\rm sc}$, $i=H, L$.
These generally differ from the corresponding ground-state deformations,
$\varepsilon_i^{\rm gs}$.
The associated distortion energies,
$E_i^{\rm dist} = M_i(\varepsilon_i^{\rm sc})-M_i(\varepsilon_i^{\rm gs})$,
are converted into statistical fragment excitations later on
as the fragment shapes relax to their ground-state forms.
(Strictly speaking, a fragment acquires the equilibrium shape
dictated by its degree of excitation, $\varepsilon_i^{\rm eq}$,
but the difference between that and $\varepsilon_i^{\rm gs}$
is small and is ignored here.)
The shape-dependent fragment masses, $M_i(\varepsilon)$, 
are calculated in the same microscopic-macroscopic model 
that was used to obtain the potential-energy surfaces \cite{MollerPRL92}.

\begin{figure}[bt]
\centering
\includegraphics[width=1.0\linewidth]{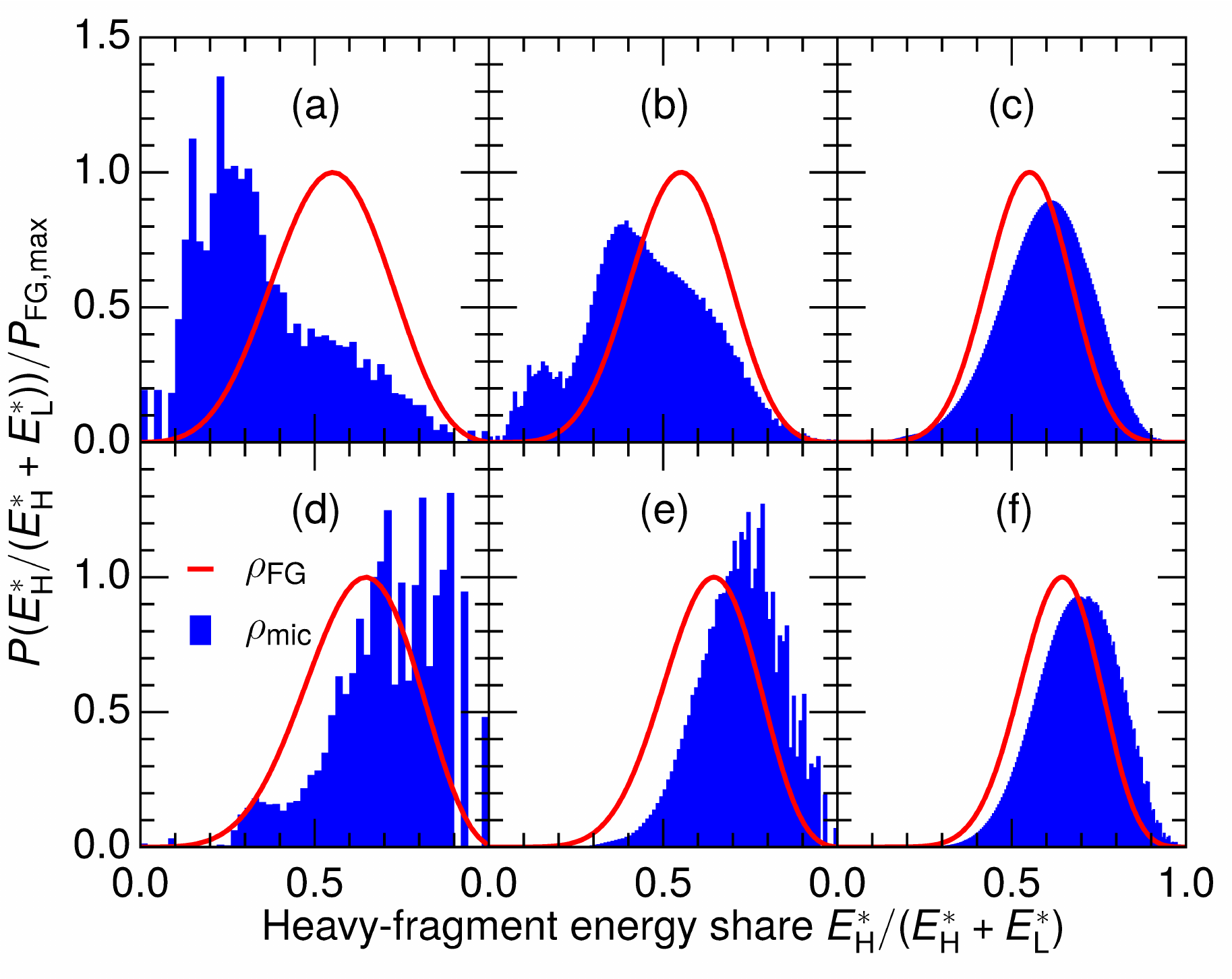}
\ccaption{\label{f1}
The distribution function $P(E_H^*;E_{\rm sc}^*)$
for the total excitation of the heavy fragment
in $^{235}$U(n,f) for two different divisions,
either $(N,Z,\varepsilon)_H = (80,50,-0.1)$
and    $(N,Z,\varepsilon)_L = (64,42,0.3)$ (top panels)
or     $(N,Z,\varepsilon)_H = (92,60,0.1)$
and    $(N,Z,\varepsilon)_L = (52,32,0.1)$ (bottom panels),
and three different values of the available energy at scission,
$E_{\rm sc}^*$=10 (left column), 20 (center column), 40 (right column) MeV. 
The distributions obtained from microscopic (blue histograms) 
and Fermi-gas (solid red curves) level densities 
are normalized to the maximum value of the Fermi-gas result.} 
\end{figure}

Because (by assumption) the collective kinetic energy associated with
the shape evolution is negligible prior to scission,
the available excitation energy at scission is 
the difference between the total energy, $E_{\rm tot}$,
and the potential energy of the scission configuration,
\begin{equation}\label{Esc}
E_{\rm sc}^*\ =\ E_{\rm tot} - U(Q_{\rm sc},c_{\rm sc},
       \varepsilon_1^{\rm sc},\varepsilon_2^{\rm sc},\alpha_{\rm sc})\ .
\end{equation}
In the present study, 
we assume that this quantity is divided statistically
between the two fragments, 
{\em i.e.}\ the total excitation energy of the heavy fragment, $E_H^*$,
is governed by the following microcanonical distribution,
\begin{equation}\label{prob}
P(E_H^*;E_{\rm sc}^*)\ \sim\
	     \tilde{\rho}_H(E_H^*;\varepsilon_H^{\rm sc})\,
	     \tilde{\rho}_L(E_{\rm sc}^*-E_H^*;\varepsilon_L^{\rm sc})\ ,
\end{equation}
and $E_L^*\!=\!E_{\rm sc}^*-E_H^*$ due to energy conservation, where 
$\tilde{\rho}_i(E_i^*;\varepsilon_i^{\rm sc})
	\equiv \tilde{\rho}(N_i,Z_i,E_i^*;\varepsilon_i)$
is the effective density of states (see below)
of a nucleus with neutron and proton numbers
$N_i$ and $Z_i$, spheroidal deformation $\varepsilon_i$, 
and a total excitation energy of $E_i^*$, with $i=H,L$.

The key novelty of the present study is the use of shape-dependent
microscopic level densities in the above expression (\ref{prob})
for the partitioning of the available energy.
The fragment level densities are calculated using the combinatorial model 
of Ref.\ \cite{UhrenholtNPA913},
using the same model as that giving the shape-dependent 
compound nuclear level density employed in the Metropolis walk.
Thus, for each emerging fragment, 
the neutron and proton wave functions are calculated
in the spheroidal effective field
and the many-quasi-particle excitations are constructed. 
For each such configuration, a pairing calculation is carried out 
and the associated rotational band is built.
For each value of the angular momentum $I$,
the level density $\rho(E^*,I;\varepsilon^{\rm sc})$ is then extracted by
counting the number of energy levels having angular momentum $I$
in a small energy interval around $E^*$.
In the present study, we are interested in the energy distribution only,
so we sum over the fragment angular momentum, $I_i$,
to obtain the effective density of states entering in Eq.\ (\ref{prob}),
\begin{equation}\label{rhoeff}
\tilde{\rho}_i(E_i^*;\varepsilon_i^{\rm sc})\ =\ 
\sum_{I_i} (2I_i+1)\, \rho_i(E_i^*,I_i;\varepsilon_i^{\rm sc})\ .
\end{equation}

Figure \ref{f1} shows the energy distribution $P(E_H;E_{\rm sc}^*)$
at three different values of the total available energy $E_{\rm sc}^*$ 
for two different mass divisions
having $(A_H$:$A_L)$ = (130:106) and (152:84).
These two divisions contribute to the yields at the inner and outer wings
of the double-humped mass distribution, respectively,
(see {\em e.g.}\ Fig.\ 9 of Ref.\ \cite{WardPRC95}), 
and the deformations considered are typical of those divisions.

\begin{figure}[tb]
\centering
\includegraphics[width=1.0\linewidth]{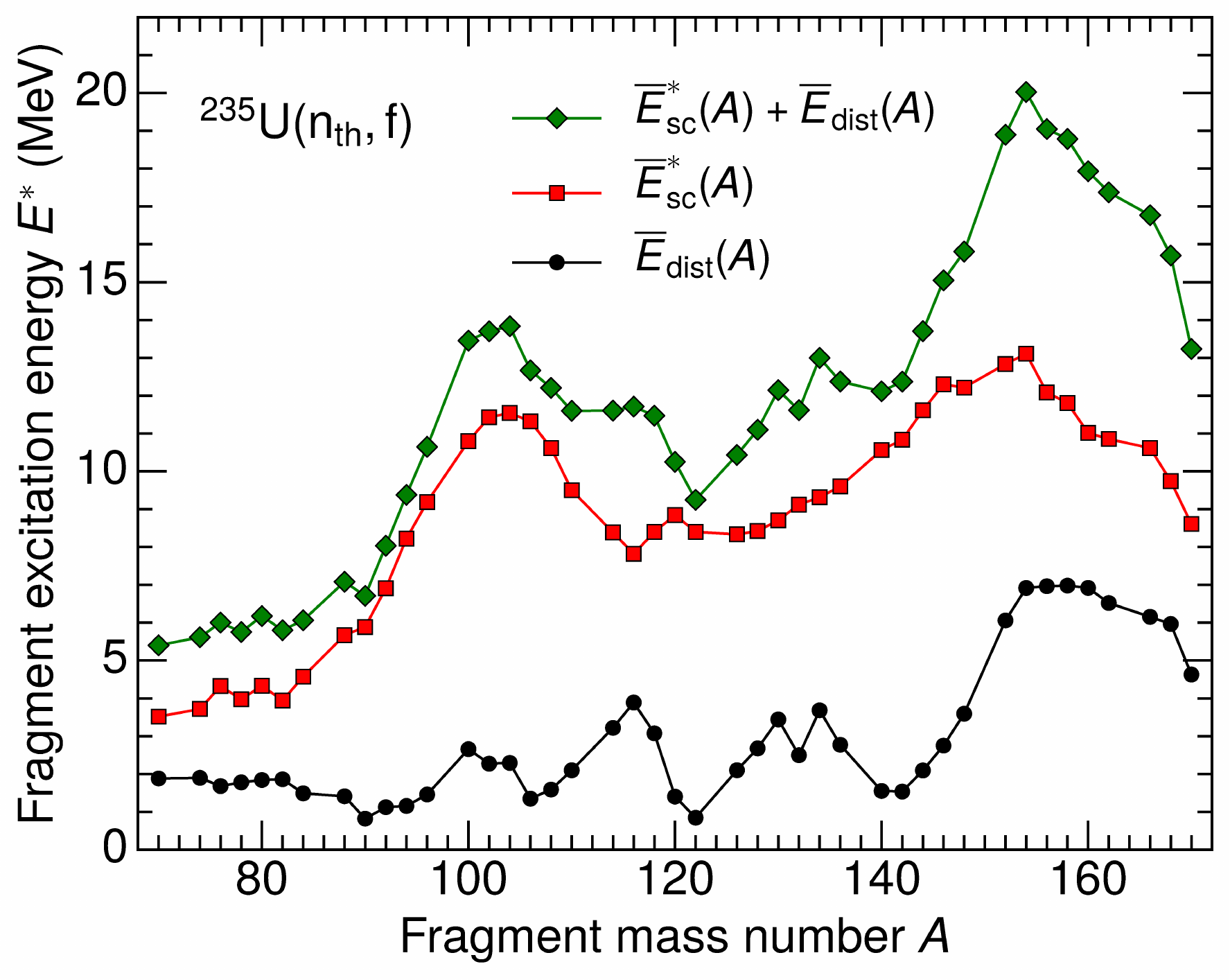}
\ccaption{\label{f2} 
\label{f2} As functions of the fragment mass number $A$ are shown
the mean fragment distortion energy $E_{\rm dist}(A)$ (black dots),
the mean excitation energy at scission, $E_{\rm sc}^*(A)$ (red squares),
and the sum, $E_{\rm dist}(A)+E_{\rm sc}^*(A)$ (green diamonds),
as extracted from an ensemble of $10^6$ scission configurations.}
\end{figure}

The energy distribution was calculated with
both the microscopic level density discussed above 
and a simplified Fermi-gas level density, 
$\rho_{\rm FG}(E^*)\sim\exp[2\sqrt{aE^*}]$
with $a$=$A/(8\,{\rm MeV})$.
Both yield rather broad distributions
due to the smallness of the nuclear system.
The macroscopic form yields smooth 
Gaussian-like distributions peaked at $E_H^*/E_L^*=A_H/A_L$,
whereas the microscopic form yields irregular distributions
that may have qualitatively different appearances,
especially at lower values of $E_{\rm sc}^*$
where quantal structure effects are most significant.
In particular, 
it is possible that one fragment receives all the available energy
with the partner fragment being left without excitation.
Although the probability for this decreases quite rapidly 
with increasing $E_{\rm sc}^*$, 
this feature is in dramatic contrast to the macrocopic result.

For the case shown in the top panel of Fig.\ \ref{f1},
the heavy fragment is $^{130}$Sn which is very close to being doubly magic.
It therefore has a spherical ground-state shape, $\varepsilon_H^{\rm gs}=0$, 
while the light fragment, $^{106}$Mo, has 
a well-deformed prolate ground-state shape, $\varepsilon_L^{\rm gs}=0.33$. 
The fragment deformations at scission are
$\varepsilon_H^{\rm sc}=-0.10$ and $\varepsilon_L^{\rm sc}=0.30$
which both deviate only slightly from the ground-state deformations.
The near magicity of the heavy fragment 
(with a shell correction energy of -10.2 MeV) 
causes the level density to remain very small up to excitation
energies of 20 MeV. 
Conversely, the shell correction energy of the light fragment is +0.35 MeV
and its level density is considerably larger than that of
the heavy partner in that energy range.
As a consequence, 
the energy distribution is peaked at small values of $E_H^*$ 
and the major part of the energy goes to the light fragment.
For example, when the total energy available for sharing is 10 MeV,
the most likely outcome is that 
the heavy fragment receives only $\approx$2 MeV,
while the light fragment gets $\approx$8 MeV. 
This is very different from the macroscopic (Fermi-gas) scenario 
in which the most likely excitations of those fragments 
are about 5.5 and 4.5 MeV, respectively.

The opposite appears when the two fragments differ more in size,
as illustrated in the bottom panel of Fig.\ \ref{f1}.
Here the microscopic energy-partition distribution function 
strongly favors the heavy fragment, $^{152}$Nd, 
relative to the light fragment, $^{84}$Ge.
In this case, the typical scission deformation of the heavy fragment 
is considerably smaller, $\varepsilon_H^{\rm sc}=0.10$, 
than its ground-state deformation, $\varepsilon_H^{\rm gs}=0.24$. 
Therefore the heavy fragment has a large single-particle level density and, 
consequently, it has a large positive shell correction, $+6.1$\,MeV
(as compared to $-6.9$\,MeV for the ground-state shape)
and a particularly high level density.
On the other hand, the neutron number of the light fragment, $N_L=52$, 
is close to being magic so its level density is low.
A a result, the heavy fragment is strongly favored in the
energy division, even up to quite high energies,
as clearly seen in Fig.\ \ref{f1} (d)-(f).

As the available energy is increased,
the microscopic energy partition distribution (\ref{prob}) approaches the 
macroscopic form obtained with the Fermi-gas level density \cite{WardPRC95}
and the structure effects on the mass partition subside,
albeit at various rates.

For each scission configuration obtained at the end of the Metropolis walk,
the excitation energies of the nascent fragments are sampled
from the appropriate microscopic partition distribution (\ref{prob})
illustrated in Fig.\ \ref{f1}.
For $^{235}$U(n$_{\rm th}$,f),
the resulting mean excitation energy $\overline{E}^*_{\rm sc}(A)$ is shown 
in Fig.\ \ref{f2} as a function of the fragment mass number $A$,
together with the mean fragment distortion energy 
$\overline{E}_{\rm dist}(A)$, 
as well as the sum of these two quantities
which represents the total excitation energy of the fragment
relative its ground state.

The mean fragment excitation energy at scission has a pronounced structure 
that may be qualitatively understood  from the energy partition distribution
functions of the two examples discussed above.
The local minimum slightly below $A=130$ 
and the local maximum around $A=106$ 
result from the favoring of the light fragment in the energy sharing
illustrated in Fig.\ \ref{f1} (a) -- (c),
while the pronounced maximum at $A\approx150$ 
and the relatively low values in the $A\approx84$ region
result from the favoring of the heavy fragment
illustrated in Fig.\ \ref{f1} (d) -- (f).

With regard to the distortion energies,
we note that the scission shapes are typically less deformed
than the corresponding ground-state shapes, 
$\varepsilon_i^{\rm sc}<\varepsilon_i^{\rm gs}$.
The resulting mean distortion energies increase from 2--3 MeV 
for light fragments to 6--7 MeV for heavy fragments.
As a consequence, the maximum in $\overline{E}^*_{\rm sc}(A)$ around $A=150$ 
is enhanced by the large distortion energies in the same mass region,
as is clearly brought out in Fig.\ \ref{f2}.

\begin{figure}[t]
\centering
\includegraphics[width=1.0\linewidth]{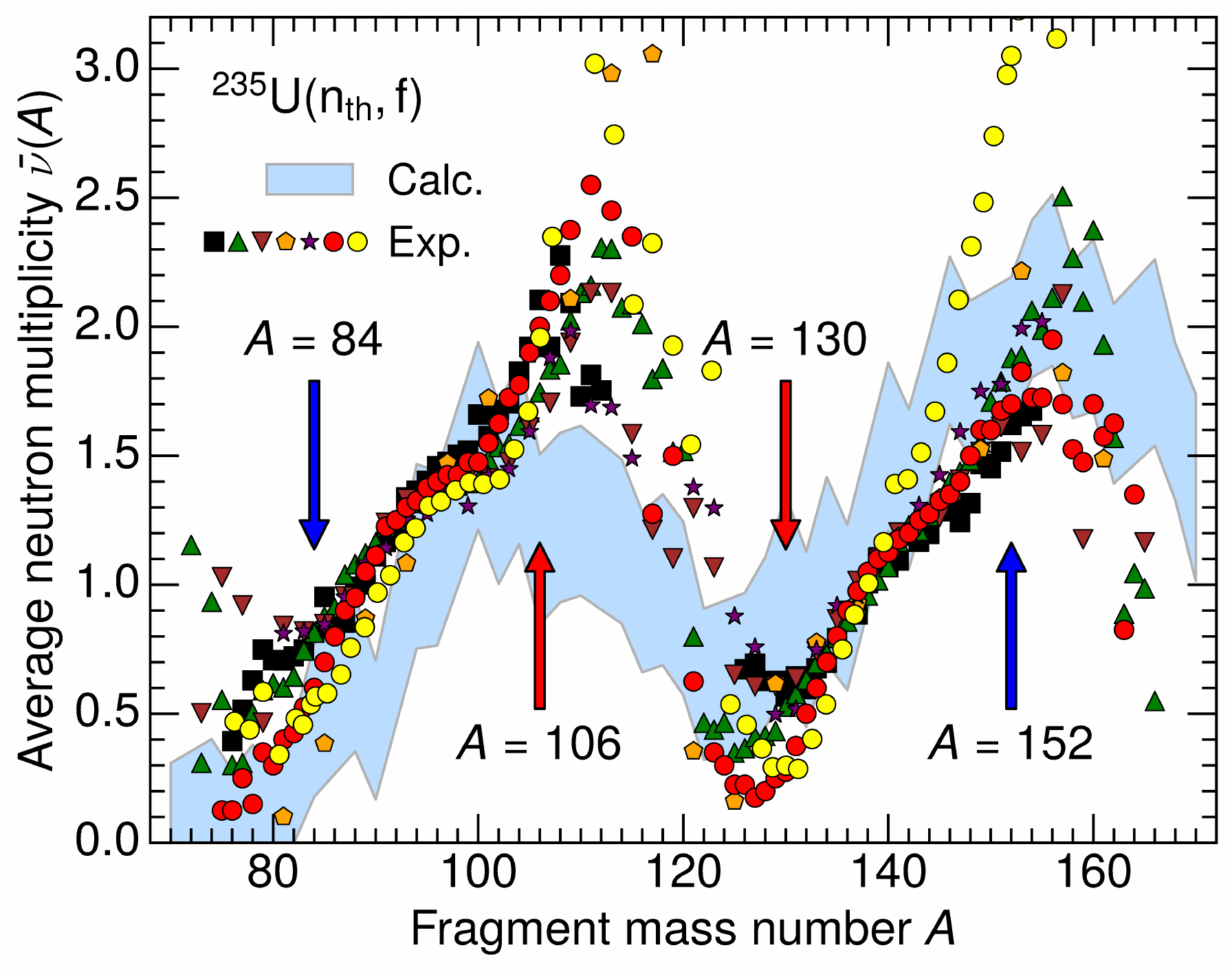}
\ccaption{\label{f3}
For $^{235}$U(n$_{\rm th}$,f) the calculated
mean neutron multiplicity as a function of the mass number
of the primary fission fragment, $\bar{\nu}(A)$, is shown 
together with a variety of experimental data:
black squares \cite{nishio236u}, 
yellow circles \cite{apalin1965},
green triangles \cite{vorobyev2010}, 
orange diamonds \cite{batenkov2005}, 
purple stars \cite{boldeman1971}
brown triangles \cite{maslin1967}, 
red circles \cite{gook2018}.
The calculated values are at the center of the light-blue shaded band
which has a width equal to the calculated dispersion of the neutron 
multiplicity distribution for that fragment mass, $\sigma_\nu(A)$.
The red/blue arrows show the locations of the mass divisions
selected in Fig.\ \ref{f1}.}
\end{figure}

After a fragment has been fully accelerated
and its shape has relaxed to its ground-state form,
it disposes of its excitation energy by neutron evaporation
and, on a longer time scale, by radiation of photons.
Because the number of neutrons emitted reflects
the degree of initial excitation, 
the dependence of the mean neutron multiplicity on the fragment mass,
$\bar{\nu}(A)$, may be used to test the calculated energy partitioning.
Therefore we consider neutron evaporation from the fragments. 

Because the initial compound excitation energies are relatively low,
neutron emission prior to (or during) fission is insignificant.
Furthermore, the fragment angular momentum $I$ is hardly affected 
by the evaporation, so the energy available for neutron evaporation
is taken as $E=E^*-\bar{E}_{\rm rot}$,
where $\bar{E}_{\rm rot}$ is the average rotational energy 
(which will later contribute to the photon radiation).
For a given fragment $(Z,N,E,\varepsilon)$,
the kinetic energy $\epsilon_{\rm n}$ of the evaporated neutron 
is sampled from the spectrum
$\sim\tilde{\rho}'(E';\varepsilon')\,\epsilon_{\rm n}$,
where $\tilde{\rho}'$ denotes the effective level density in the daughter
fragment $(Z'=Z,N'=N-1,E'=E-\epsilon_{\rm n}-S_{\rm n},\varepsilon')$,
with $S_{\rm n}$ being the neutron separation energy in the mother fragment.
For consistency, we employ the microscopic level density (\ref{rhoeff}) 
for the evaporation daughter nucleus.
Following the treatment in Ref.\ \cite{RandrupPRC80},
the neutron evaporation is continued until the excitation energy 
has fallen below the neutron separation energy.

\begin{figure}[t]
\centering
\includegraphics[width=1.0\linewidth]{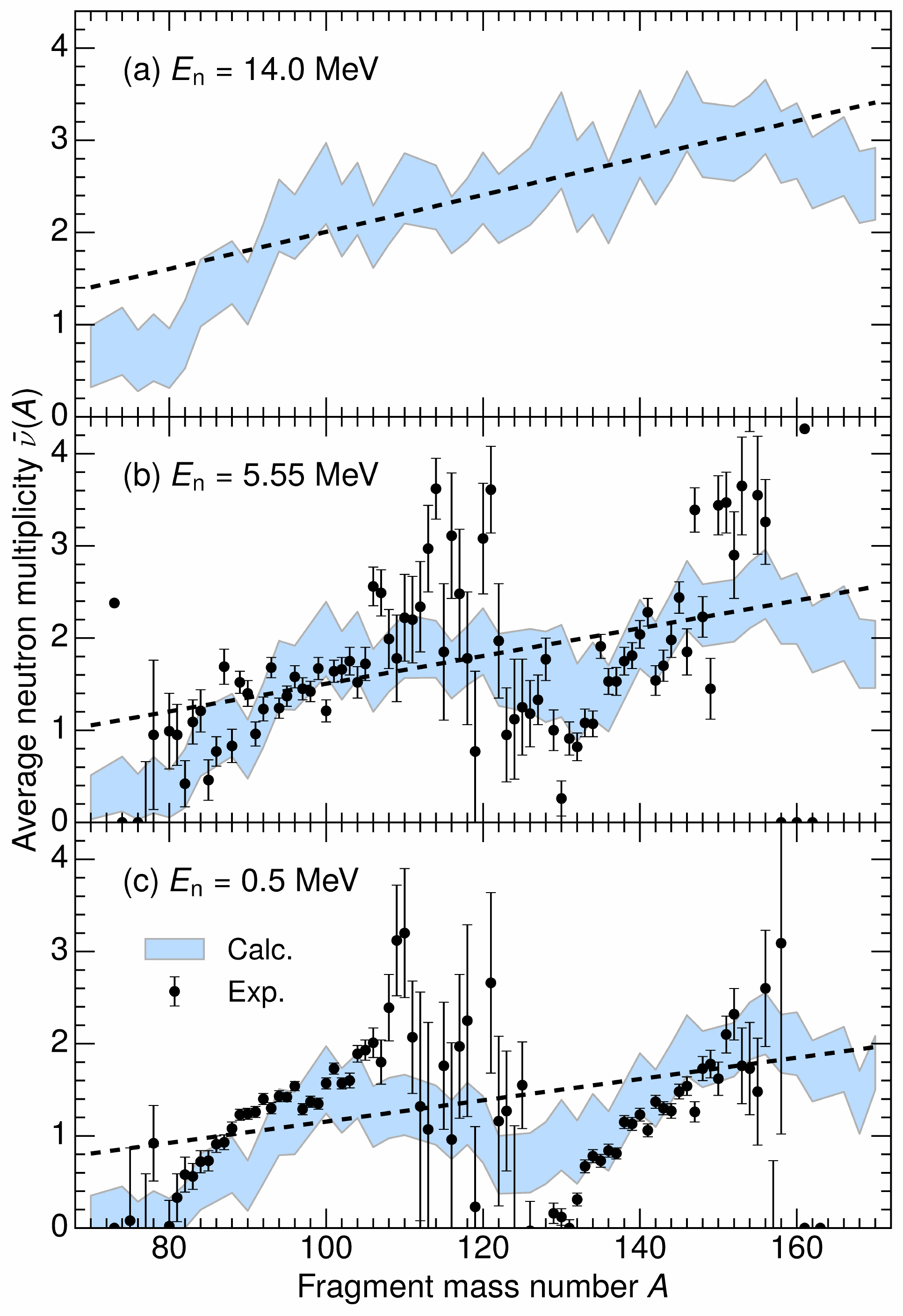}
\ccaption{\label{f4} 
For $^{235}$U(n,f) is shown the mean neutron multiplicity 
as a function of the mass number of the primary fission fragment,
$\bar{\nu}(A)$,	
for three different incident neutron energies $E_{\rm n}$:
0.5 MeV (a), 5.55 MeV (b), 14 MeV (c).
The calculated values are at the center of the light-blue shaded band
which has a width equal to the calculated dispersion of the neutron 
multiplicity distribution for that fragment mass, $\sigma_\nu(A)$.
The experimental data from Ref.\ \cite{muller1984} are also shown.
The dashed line shows roughly the behavior
resulting from an energy division according to mass.}
\end{figure}

Figure \ref{f3} shows the calculated mean neutron multiplicity $\bar{\nu}(A)$ 
together with experimental data from a variety of experiments.
The calculated $\bar{\nu}(A)$ is at the center of the light-blue band 
which has a width equal to the dispersion of the 
calculated neutron multiplicity distribution for that $A$, $\sigma_\nu(A)$.
The band is shown to make it easier to identify
the calculated value on the plot and, at the same time, 
to give a quantitative impression of the fluctuation 
in the number of neutrons emitted from a fragment.

The sawtooth appearance of the data is reasonably well reproduced 
by the calculation and arises from a combined effect 
of the behavior of the neutron separation energy $S_{\rm n}(A)$, 
which displays a jump near $A=132$ due to the closed shells
at $Z=50$ and $N=82$, and the behavior of the total intrinsic fragment energy
$E_{\rm dist}(A)+E_{\rm sc}^*(A)$ (see Fig.\ \ref{f2}).

The energy dependence of the energy partitioning 
is illustrated in Fig.\ \ref{f4}
which shows $\bar{\nu}(A)$ resulting from first-chance fission
at three different incident neutron energies.
The experimental data from Ref.\ \cite{muller1984} are also shown.
The dashed line shows roughly the neutron multiplicity
resulting if the excitation energy were divided according to the masses
as suggested by the simple Fermi-gas level density.
It is seen that the calculated results approach this behavior
with increasing excitation energy.
In the region around $A=130$,
the very low neutron multiplicity occuring for thermal fission 
grows rather rapidly with increasing neutron energy,
causing the sawtooth feature of $\bar{\nu}(A)$ to become smoother.
This behavior is due to the decrease of the strong negative shell correction 
at higher excitation energy for fragments in this mass region, 
increasing the level density and thus the share of the excitation energy
taken up by the heavy fragment at scission.

In summary, in order to elucidate how the available excitation energy 
at scission is divided between the two fragments, 
we have augmented the recently developed level-density guided
Metropolis shape evolution treatment \cite{WardPRC95}
with shape-dependent microscopic level densities 
for the nascent fission fragments
which are distorted relative to their equilibrium shapes.
The available energy is partitioned statistically
according to the corresponding microscopic level densities
which take account of the structure effects in these distorted pre-fragments.
For each fragment, 
the distortion energy is converted into additional excitation
before they experience sequential neutron evaporation.
The dependence of the resulting mean neutron multiplicity on fragment mass,
$\overline{\nu}(A)$, agrees well with experimental data.
In particular, the sawtooth appearance of $\overline{\nu}(A)$ 
can be understood from shell-structure effects in the level densities 
as well as from structure in the deformation energy surface.

We also studied how $\overline{\nu}(A)$ changes 
as the excitation energy of the fissioning nucleus is increased.
The sawtooth behaviour is weakened due to the reduction 
of the shell corrections near $A=130$
which significantly increases the level density in the heavy fragment
and hence $\bar{\nu}$.
Such an evolution is also seen in the experimental data.

It is notewhorty that the presented treatment
stays within the well-established framework of the 
macroscopic-microscopic model of nuclear structure
underlying the calculation of the nuclear potential-energy surfaces
that have been used successfully to calculate 
fission-fragment mass distributions
\cite{RandrupPRL106,RandrupPRC84,WardPRC95,RandrupPRC88}.
This novel treatment has considerable predictive power
and can readily be applied to other fission cases as well,
including cases where no experimental data yet exist.\\[-1ex]

This work was supported by 
the Swedish Natural Science Research Council (S.\AA.) 
and the Knut and Alice Wallenberg Foundation (M.A., B.G.C. and S.\AA.); 
J.R.\ was supported in part by the NNSA DNN R{\&}D 
of the U.S.\ Department of Energy.

\end{document}